\documentclass{article}
\usepackage{spconf}
\usepackage{graphicx}
\usepackage[inkscapelatex=false]{svg}
\usepackage{booktabs}
\usepackage{amsmath,amssymb,bm}
\usepackage{textcomp}
\usepackage{xcolor}
\usepackage{hyperref}

\begin{document}
\ninept
\makeatletter
\renewcommand{\footnotesize}{\@setfontsize{\footnotesize}{9pt}{10.5pt}}
\makeatother
\title{Randomized Scores and Diverse Timbres: Augmenting Automatic Music Transcription with Online-Generated Data}

\name{Haiwen Xia$^1$, Chao Zhang$^2$, Qiuqiang Kong$^3$\sthanks{Corresponding author: Qiuqiang Kong.}}
\address{
  \begin{tabular}{@{}c@{\hspace{1.5em}}c@{\hspace{1.5em}}c@{}}
    $^{1}$Peking University                         &
    $^{2}$Tsinghua University                       &
    $^{3}$The Chinese University of Hong Kong         \\
    {\normalsize\texttt{xiahaiwenm@stu.pku.edu.cn}} &
    {\normalsize\texttt{cz277@tsinghua.edu.cn}}     &
    {\normalsize\texttt{qqkong@ee.cuhk.edu.hk}}
  \end{tabular}
}

\maketitle
\begin{abstract}
  Automatic music transcription (AMT) is limited by the scarcity of audio
  recordings paired with precise symbolic annotations. Synthetic data can
  provide supervision at scale, but it remains unclear whether effective
  transfer depends on realistic score structure or broad timbral coverage.
  We study these factors separately through an online \emph{sampler--renderer}
  pipeline. A unified corruption sampler ranges from unmodified MIDI clips
  through partial corruption to deeply randomized note-event distributions.
  The renderer converts these events to audio while independently controlling
  instrument and timbral coverage. A fixed transcription model is trained jointly on
  offline recordings and newly rendered examples. Controlled ablations reveal
  an asymmetry between the two factors: moderate corruption of the note-event
  distribution does not impair transfer and can improve it, whereas broader
  renderer-side timbral
  support consistently improves out-of-domain generalization under a fixed
  note-event distribution. Finally, online-rendered examples complement real
  and existing synthetic data in a strong combined-data regime. These results
  suggest that synthetic AMT data should prioritize coverage of note-level
  attributes and their timbral realizations over realistic joint score
  structure.
\end{abstract}


\section{Introduction}
\label{sec:introduction}
Automatic music transcription (AMT) converts recordings into symbolic note
representations~\cite{benetos2018automatic}, supporting tasks music education, automatic music understanding and controllable generation. Approaches have progressed from non-negative matrix factorization
(NMF)~\cite{smaragdis2003nmf} through deep neural networks
(DNNs)~\cite{sigtia2016end,kelz2016potential} and convolutional recurrent
networks (CRNNs)~\cite{onsets,kong2021high} to Transformer-based models
~\cite{hawthorne2021sequence,MT3,PercieverTF}.Strong multi-instrument benchmark results have been achieved by adapting
language-model architectures such as YourMT3+~\cite{Chang2024YourMT3+} and MuScriptor~\cite{Muscriptor_2026}, yet
transcription in the wild remains difficult~\cite{MulTTiPop}.

A key obstacle is the scarcity of diverse recordings paired with precise note annotations. Annotating precise polyphonic notes is costly, leaving
supervised datasets with limited coverage of musical and acoustic conditions.
Due to the lack of paired audio--symbolic data, synthetic supervision offers
a practical alternative: symbolic events provide labels for rendered audio,
allowing training data to expand without manually annotating each recording.

Existing methods follow two complementary strategies. Rendering-based methods synthesize audio from symbolic scores using sampled or modeled instruments \cite{Muscriptor_2026,Slakh,EdwardsDBMK24AcousticAug,Zang2024SynthTab}, expanding acoustic coverage while retaining coherent source scores. Closely related to our note renderer, Sato and Akama \cite{Sato2024NsynthAug} combine Lakh scores with isolated NSynth notes \cite{engel2017nsynth}. Complementing these rendering approaches, augmentation-based
methods recombine existing audio--label pairs, such as monophonic tracks, stems, or drum
patterns~\cite{Chang2024YourMT3+,Simon2022mixing,egmd}. Their success suggests that realistic arrangements can be
relaxed, but recombination still preserves structure within each source
and changes polyphony and timbre together. It therefore
remains unclear which score dependencies are necessary and how timbral coverage
should be varied independently.

These observations motivate our online sampler--renderer formulation. A symbolic sampler first selects note events, ranging from source clips to
corrupted or note-factorized attributes. A renderer then converts these
events into audio using selected timbres, with the sampled
events supplying the transcription targets. Holding
one fixed while varying the other isolates score structure from timbral
coverage. Each pair is generated once during training, exposing new
event--timbre combinations alongside fixed offline recordings and enabling
controlled transfer comparisons.

Our contributions are fourfold: \textbf{First}, we show that moderately
randomizing note events preserves and can improve transfer, so realistic score
structure is largely unnecessary. \textbf{Second}, we show that broadening
timbral coverage under a fixed note-event distribution gives monotonic
out-of-domain gains. \textbf{Third}, we develop a simplified multi-channel
architecture that splits instrument channels after encoder projection and
shares decoder weights without a specialized Perceiver~TF encoder.
\textbf{Fourth}, we show that online-rendered data remain complementary to real
and existing synthetic datasets when all sources are trained jointly.
Our code is open-sourced at
  {\footnotesize\url{https://github.com/Haiwen-Xia/randomscore-amt}} to facilitate AMT research.

The remainder of the paper is organized as follows. Section~II presents the
online sampler--renderer formulation. Section~III describes our model architecture. Section~IV reports the experiments
and limitations, and Section~V concludes the paper.


\section{Online Sampler--Renderer Data Generation}
\label{sec:method}

\subsection{Problem Formulation}

Let $\bm{x} \in \mathbb{R}^L$ denote an audio segment and $\bm{y}$ its note events. For this work, note events refer to note-oriented MIDI before tokenization.
During training, we synthesize labeled \textit{online} pairs: each pair is
rendered on the fly and used once, unlike \textit{offline} pairs, which are
fixed and revisited every epoch. Generation factorizes into two
independent controls: a \textit{sampler} determines the note events
$\bm{y}$, and a \textit{renderer} determines their acoustic
realization $\bm{x}$. Fixing one while varying the other isolates
score structure from timbral coverage. Section~II-B defines the
\textit{sampler} and Section~II-C introduces the \textit{renderer}.

\begin{figure}[t]
  \centering
  \includegraphics[width=\linewidth]{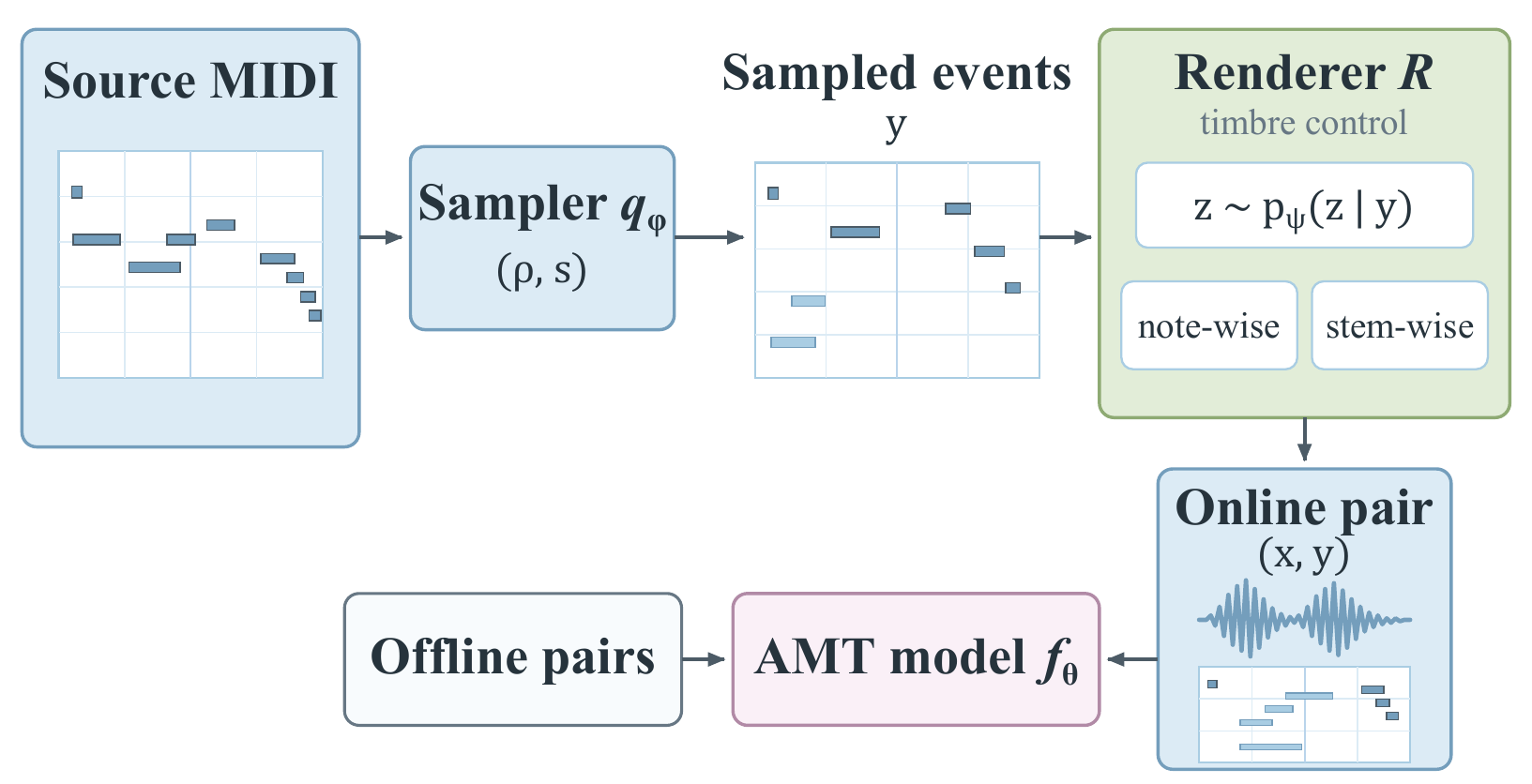}
  \caption{Online data generation with independent controls over score
    structure and timbre, combined with offline data for joint training.}
  \label{fig:pipeline}
\end{figure}

\subsection{Symbolic Sampler}
The symbolic sampler produces the target note set $\bm y=\{\bm e_n'\}_{n=1}^{N}$
by corrupting a sampled MIDI clip with notes $\bm e_1,\ldots,\bm e_N$, each
carrying the attributes (program, pitch, onset, offset, velocity). We write
$q_\phi(\bm y)$ for the resulting distribution, parameterized by
$(\rho,s)$ below.

We define
a factorized note distribution $r(\bm e)$, which treats notes as independent: it retains each note's program, draws pitch and duration from
their empirical marginals estimated on 2.048-s segments of the source corpus. It
draws onset and velocity uniformly over
their feasible ranges. We avoid uniform duration sampling because excessive
long notes would make the transcription problem ill-posed.

Each clip is assigned a corruption rate $\rho\in[0,1]$, i.e., the fraction
of notes to be altered. Every note is then independently replaced by an
intermediate note
\begin{equation}
  \widetilde{\bm e}_n=
  \begin{cases}
    \bm e_n, & \text{with probability }1-\rho, \\
    \widetilde{\bm e},\quad \widetilde{\bm e}\sim r(\bm e),
             & \text{with probability }\rho.
  \end{cases}
  \label{eq:corruption-draw}
\end{equation}
The output note interpolates between the original and intermediate notes, with an interpolation strength $s$:
\begin{equation}
  \bm e_n'=(1-s)\bm e_n+s\widetilde{\bm e}_n,
  \qquad s\in[0,1],
  \label{eq:corruption-interpolation}
\end{equation}
Because r(e) retains the original program, interpolation leaves it unchanged.
$s=0$ keeps the original note and
$s=1$ is full replacement. Pitch is rounded to a valid MIDI
value after interpolation. $\rho$ controls how many notes are altered, while $s$ controls
how far each altered note moves from its original value.

The parameters $(\rho,s)$ select the three settings that we compare:
\begin{itemize}\setlength{\itemsep}{0pt}
  \item \emph{Source} ($\rho=0$): the sampled clip is returned unchanged;
  \item \emph{Corrupted} ($\rho\sim\mathcal U[0,1]$ drawn per clip, $s>0$):
        selected notes move toward their factorized draws;
  \item \emph{Note-factorized} ($\rho=1,s=1$): every note's non-program attributes are
        redrawn independently.
\end{itemize}
The three settings are therefore instances of a single sampler rather than
separate algorithms.

\subsection{Audio Renderer and Timbral Coverage}

Given a symbolic target $\bm y$ from Section~II-B, the renderer samples a
compatible timbral choice $\bm z\sim p_\psi(\bm z\mid\bm y)$ and produces audio \quad $ \bm x=R(\bm y,\bm z)$:
Here, $\bm x$ is the rendered audio and $\bm z$ is the concrete renderer
choice: compatible NSynth recordings for note-wise rendering or a Pianoteq
preset for stem rendering. Broadening the support of these choices increases
timbral coverage without changing the symbolic distribution.

\subsubsection{Note-renderer}
The note-wise renderer maps each symbolic note to a compatible NSynth
recording~\cite{engel2017nsynth} and mixes the rendered notes at their target
onsets. Directly clipping a recording can misalign its sounding duration with
the symbolic offset. We therefore detect the sounding region and align its
start to the target onset, so that each note is audible from its onset, and
apply a smooth decay when shortening a note. This duration mapping
uses local per-note operations, allowing notes to be processed in parallel with
low overhead.

\subsubsection{Stem renderer}
The stem renderer sends a complete MIDI stem to
Pianoteq\footnote{Pianoteq 9.1.2 free version, \url{https://www.modartt.com/}. The free version excludes 8 fixed pitches (F$\sharp$1, G$\sharp$1, A$\sharp$1, C$\sharp$5, D$\sharp$5, F$\sharp$5, G$\sharp$5, A$\sharp$5).} and samples one
compatible preset. The free version exposes 797 presets across 25 instrument
groups but excludes 8 fixed pitches; affected notes are relabeled to the
nearest playable octave of the same pitch class.
Although our pipeline controls only the preset, each preset configures coupled
physical-model parameters such as unison width, sympathetic resonance, and
hammer hardness. Preset sampling therefore changes multiple aspects of timbre
while retaining a coherent realization across all notes in a stem. To amortize
engine startup, we concatenate multiple MIDI clips in one invocation and split
the rendered waveform at known clip boundaries. For controlled piano
experiments, we compare subsets containing 4\%, 20\%, and 100\% of program-0 (Acoustic Grand Piano)
presets, followed by presets spanning programs 0--7 (Broad Piano).

The rendered audio $\bm x$ and its exact symbolic target $\bm y$ form one
online pair. Each minibatch uses a fixed offline-to-online ratio, with all
samples the same loss weight. We apply the same intra-stem dropping and cross-stem
mixing augmentation in every condition.
\section{Autoregressive Transcription Model}
\label{sec:model}
\begin{figure}[!t]
  \centering
  \includegraphics[width=\linewidth]{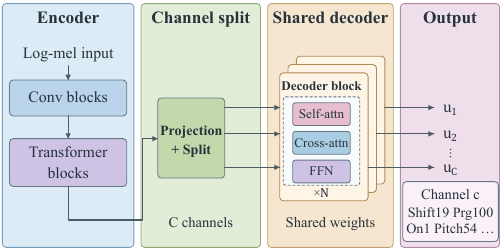}
  \caption{Transcription architecture with shared decoder weights across channels.}
  \label{fig:arch}
\end{figure}

We transcribe with a compact encoder--decoder: a convolutional front end and a
temporal Transformer encoder summarize the spectrogram into a shared acoustic
representation, which is projected and split into $C$ channel streams, one per
coarse instrument class ($C{=}13$, following YourMT3+~\cite{Chang2024YourMT3+});
each stream is then decoded autoregressively into event
tokens by a decoder whose parameters are shared across all channels
(Fig.~\ref{fig:arch}). Unlike YourMT3+, we replace its specialized Perceiver
TF~\cite{PercieverTF} encoder with a simpler stack. The model is held fixed
within each data-generation comparison.

\subsection{Encoder and Channel Split}
Let $X\in\mathbb{R}^{T\times F}$ be the log-mel spectrogram of one clip, with
$T$ mel frames and $F$ mel bins. Convolutional blocks filter time and frequency
and pool only along frequency, preserving temporal resolution; temporal
Transformer blocks then model longer-range context and yield shared features
$E\in\mathbb{R}^{T\times D}$, where $D$ is the encoder width. A linear
projection expands every frame of $E$ to width $Cd$ and splits the result into
$C$ streams $H_1,\dots,H_C\in\mathbb{R}^{T\times d}$ of equal width $d$. All
Transformer blocks use rotary position embeddings (RoPE) and
RMSNorm~\cite{su2024roformer,zhang2019rmsnorm}.

\subsection{Multi-Channel Decoder}
Let $M_c$ be the number of event tokens in channel $c$, with $u_{c,m}$ the
$m$-th token and $u_{c,<m}$ its prefix. Given the audio, channels are
conditionally independent and decoded in parallel, while tokens within a
channel remain autoregressively dependent on their own prefix using RoPE. Each decoder
block applies causal self-attention over the channel's own prefix,
cross-attention to the corresponding encoder stream $H_c$, and a position-wise
feed-forward transformation. Token embeddings, block parameters, and the output
head are shared by all $C$ channels, so channels differ only in the stream they
attend to and in the events they emit. Event tokens encode instrument, pitch, event time and note on-off over
a vocabulary of size $V$; in piano-only experiments only the piano channel is
active.

\subsection{Training Objective}
This factorization yields the autoregressive token cross entropy
\begin{equation}
  \mathcal{L}_{\mathrm{AR}}=-\sum_{c=1}^{C}\sum_{m=1}^{M_c}
  \log p_\theta(u_{c,m}\mid u_{c,<m},H_c).
  \label{eq:ar-loss}
\end{equation}
Online and offline examples share this objective and differ only in
provenance, which sets batch composition (Section~\ref{sec:method}) and not the
loss. For sources without an exact program label, probability mass is
aggregated over the corresponding instrument family.
\section{Experiments}
\label{sec:experiments}

\subsection{Experimental Setup}
All experiments share the architecture and differ in
training data: Experiments 1 and 2 are \textit{online-only}, ablating symbolic sampling
and timbral coverage with the model fixed, while Experiments 3 and 4 combine \textit{offline} audio
with \textit{online} data to test complementarity.

Following YourMT3+~\cite{Chang2024YourMT3+}, inputs are 2.048-s, 16-kHz clips
represented by 128 mel bins at an 8-ms hop ($T=256$). We use Adam with 1,000
warmup steps, cosine learning-rate decay, and a base learning rate of
$10^{-4}$. Experiments 1--3 use Base ($D{=}768$, $d{=}192$, $6{+}6$ blocks,
$\approx$60M parameters) for 200k steps (batch 32; 8 in Experiment 2), and
Experiment 4 uses Large ($D{=}1024$, $d{=}384$, $10{+}6$ blocks,
$\approx$190M) for 320k steps.

All experiments evaluate on MAPS~\cite{maps} for out-of-domain piano
transfer; Experiments 1, 3, and 4 also on the real multi-instrument recordings
MulTTiPop~\cite{MulTTiPop} and RWC~\cite{rwc}; Experiments 1 and 3 on
Slakh/test, and Experiment 2 on MAESTRO/test~\cite{maestro}.

Experiments 3 and 4 use two online configurations: \textit{NSynth} pairs the
uniform corruption sampler ($\rho\sim\mathcal U[0,1]$, $s=1$) with the NSynth
note renderer, and \textit{Mixed} uses the same sampler with NSynth and
Pianoteq renderings in a 3:1 ratio, where the Pianoteq presets span keyboard,
guitar, chromatic percussion, and bass instruments. Source pools and the in-domain status of
each test set are stated per experiment.

We report onset $F_1$ (\%) at a 50-ms tolerance, either instrument-agnostic
(Agn) or program-aware (Prog); drums are excluded as neither renderer produces
them. MAPS is Agn only; RWC averages its five subsets.

\subsection{Randomized Scores Remain Informative}
Experiment 1 isolates symbolic sampling by fixing the NSynth renderer and
drawing every source clip from Slakh/train. The empirical marginals used by the
sampler are also estimated exclusively from 2.048-s Slakh/train segments. We
compare \textit{Source}, \textit{Corrupted}
($\rho\sim\mathcal U[0,1]$, $s\in\{0.5,1\}$), and
\textit{Note-factorized} ($\rho=1,s=1$) sampling. All conditions therefore share
the renderer and source corpus but retain different amounts of inter-note
dependence.

Table~\ref{tab:score} shows that \textit{Note-factorized} sampling remains
informative despite removing most inter-note dependencies. Although weaker
than \textit{Source}, it transfers beyond its rendered training distribution.
Both \textit{Corrupted} settings remain close to \textit{Source} on the
multi-instrument sets and improve MAPS: $s=0.5$ performs best on RWC and Slakh,
while $s=1$ performs best on MAPS. \textit{Source} remains best on MulTTiPop, possibly because MulTTiPop and the
Slakh training clips share the same Lakh MIDI origin. Thus, realistic score
structure is not required for useful supervision, while sampling a range of
corruption levels is more effective than fully factorizing non-program
attributes.

\begin{table}[t]
  \centering
  \caption{Symbolic sampler comparison with a fixed NSynth renderer:
    instrument-agnostic (Agn) onset $F_1$ (\%). $\rho$ and $s$ follow the uniform
    corruption sampler of Section~\ref{sec:method}. Best scores are bold.}
  \label{tab:score}
  {\small
    \setlength{\tabcolsep}{3pt}%
    \begin{tabular}{lcccccc}
      \toprule
      Sampler                  & $\rho$             & $s$ & MAPS           & MulTTiPop      & RWC            & Slakh          \\
      \midrule
      \textit{Source}          & 0                  & 0   & 64.99          & \textbf{22.92} & 36.41          & 49.55          \\
      \textit{Corrupted}       & $\mathcal{U}[0,1]$ & 0.5 & 67.28          & 21.58          & \textbf{36.70} & \textbf{50.30} \\
      \textit{Corrupted}       & $\mathcal{U}[0,1]$ & 1   & \textbf{68.28} & 21.53          & 35.72          & 49.78          \\
      \textit{Note-factorized} & 1                  & 1   & 61.18          & 12.56          & 28.21          & 36.83          \\
      \bottomrule
    \end{tabular}}
\end{table}

\subsection{Timbral Coverage Improves Transfer}
For its rendered conditions, Experiment 2 holds the symbolic clips fixed while
expanding Pianoteq preset coverage. Fixed-seed random subsets of 11, 57, and
all 285 Grand Acoustic Piano presets (4\%, 20\%, and 100\%) test within-program diversity;
programs 0--7 extend coverage to related keyboard instruments. The Pianoteq
catalog contains no Yamaha Disklaviers, the pianos on which MAESTRO and MAPS
were recorded, so every rendered condition is timbrally out-of-domain for both
test sets. Recorded
MAESTRO audio provides a reference condition. Eval $F_1$ also excludes the
8 unsupported pitches of Section~II-C, since the model never sees them
during training.
Table~\ref{tab:timbre} shows monotonic gains on both datasets as coverage
increases. Extending to programs 0--7 further improves MAESTRO and MAPS by
1.80 and 2.43 points, respectively, suggesting that related timbres also help
acoustic-piano transcription. This setting exceeds real-MAESTRO training on
MAPS (78.13 vs.\ 77.14), while real audio remains strongest on MAESTRO due to in-distribution training.

\begin{table}[t]
  \centering
  \caption{Pianoteq timbral coverage with fixed MAESTRO clip sampling and
    training budget: Agn onset $F_1$ (\%).}
  \label{tab:timbre}
  {\small
    \setlength{\tabcolsep}{4pt}%
    \begin{tabular}{llcc}
      \toprule
      Source   & Timbral coverage           & MAESTRO        & MAPS           \\
      \midrule
      Rendered & Pianoteq, program 0, 4\%   & 60.64          & 59.40          \\
      Rendered & Pianoteq, program 0, 20\%  & 71.13          & 68.73          \\
      Rendered & Pianoteq, program 0, 100\% & 79.25          & 75.70          \\
      Rendered & Pianoteq, programs 0--7    & 81.05          & \textbf{78.13} \\
      Recorded & MAESTRO audio              & \textbf{94.87} & 77.14          \\
      \bottomrule
    \end{tabular}}
\end{table}

\subsection{Combining Offline and Online Supervision}
Experiment 3 compares four strategies: \emph{offline} (Slakh audio),
\emph{online} (the \textit{NSynth} configuration of Section~IV-A), \emph{joint}
(16 examples from each source per batch), and \emph{two-stage} (100k online
steps, then 100k Slakh steps at learning rate $10^{-5}$ instead of $10^{-4}$).

\begin{table}[t]
  \centering
  \caption{Experiment 3: onset $F_1$ (\%) of the four training strategies above
    with the Base model. Agn/Prog denotes instrument-agnostic/program-aware. Best in each column is bold.}
  \label{tab:multi}
  {\small
    \setlength{\tabcolsep}{3pt}%
    \begin{tabular}{lccccccc}
      \toprule
      Strategy  & MAPS                    & \multicolumn{2}{c}{Slakh} & \multicolumn{2}{c}{MulTTiPop}
                & \multicolumn{2}{c}{RWC}                                                                                                                                 \\
      \cmidrule(lr){3-4}\cmidrule(lr){5-6}\cmidrule(lr){7-8}
                & Agn                     & Agn                       & Prog                          & Agn            & Prog           & Agn            & Prog           \\
      \midrule
      Offline   & 68.50                   & 75.84                     & 73.59                         & 20.84          & 12.96          & 35.94          & 30.20          \\
      Online    & 68.28                   & 49.78                     & 39.37                         & \textbf{21.53} & 13.23          & 35.72          & 28.60          \\
      Joint     & 70.86                   & 75.73                     & 73.07                         & 21.44          & \textbf{13.38} & \textbf{39.09} & \textbf{33.27} \\
      Two-stage & \textbf{71.63}          & \textbf{76.21}            & \textbf{73.70}                & 20.58          & 12.82          & 37.93          & 32.30          \\
      \bottomrule
    \end{tabular}}
\end{table}

Despite a 26-point gap on Slakh itself, offline and online training perform
similarly on the three out-of-domain sets, all within one point. The synthetic
distribution thus transfers far better than its in-domain performance alone
would suggest.

Combining the two sources improves transfer while largely preserving Slakh
performance: joint training yields the best out-of-domain Prog scores and RWC,
and two-stage the best MAPS. Although NSynth labels carry noisy program
groups---synth leads, pads, and electric organs do not correspond to real
instrument families---online training alone already matches offline Slakh on
out-of-domain Prog, and its program-aware gap appears only in-domain. Online
supervision thus provides complementary acoustic coverage even under noisy
labels, whereas accurate offline labels matter mainly for in-domain instrument
assignment.

\subsection{Strong Offline--Online Regime}
Experiment 4 discusses whether online generation remains useful with a stronger offline training mixture comprising Slakh2100, MAESTRO,
GuitarSet, URMP, MusicNet, SMT-Bass, ENST-Drums, and E-GMD. We compare three
regimes: offline only, offline with \textit{NSynth}, and offline with
\textit{Mixed}. The two online regimes mix 16 online and 16 offline examples
per batch, as in Experiment 3, and draw source clips from the training
datasets. We also list two pretrained
baselines, YourMT3+~\cite{Chang2024YourMT3+} and
Muscriptor-Large~\cite{Muscriptor_2026}, on the same evaluation sets.
Table~\ref{tab:strong}
reports the evaluation results on MAPS, MulTTiPop, and RWC tracks.

\begin{table}[t]
  \centering
  \caption{Experiment 4: onset $F_1$ (\%) after 320k steps, using the same Agn
    and Prog metrics as Table~\ref{tab:multi}. The last two rows are pretrained
    baselines and are ranked together with our regimes. Best in each column,
    baseline included, is bold.}
  \label{tab:strong}
  {\small
    \setlength{\tabcolsep}{2.8pt}%
    \begin{tabular}{lccccc}
      \toprule
      Regime         & MAPS           & \multicolumn{2}{c}{MulTTiPop} & \multicolumn{2}{c}{RWC}                                   \\
      \cmidrule(lr){3-4}\cmidrule(lr){5-6}
                     & Agn            & Agn                           & Prog                    & Agn            & Prog           \\
      \midrule
      Offline        & 86.77          & 23.28                         & 13.39                   & 44.97          & 39.20          \\
      Offline+NSynth & 86.56          & 24.04                         & 14.74                   & 44.92          & 39.03          \\
      Offline+Mixed  & \textbf{87.18} & 22.85                         & 13.26                   & 46.29          & 40.16          \\
      \midrule
      YourMT3+       & 85.83          & 19.15                         & 12.39                   & 48.39          & \textbf{43.78} \\
      MuScriptor-L   & 85.82          & \textbf{39.06}                & \textbf{28.28}          & \textbf{51.90} & 40.90          \\
      \bottomrule
    \end{tabular}}
\end{table}

Within our three regimes, online supervision remains beneficial but is strongly domain dependent. \textit{+NSynth} gives the highest MulTTiPop Agn score, improving over offline training by 0.76 points, whereas \textit{+Mixed} improves MAPS by 0.41 points and RWC by 1.32/0.96 points under the Agn/Prog metrics. However, \textit{+Mixed} reduces both MulTTiPop scores. This decrease trades recall for precision: Agn precision rises from 31.30\% to 34.02\%, while recall falls from 21.55\% to 19.78\%. Thus, adding online data does not act as uniform regularization; different renderers shift the model toward different acoustic domains.

The RWC subsets show that the \textit{+Mixed} gain is not driven by a single genre: Agn onset $F_1$ improves by 1.07--1.80 points across all five subsets, and Prog improves in four of them ($-0.74$ to $+2.48$ points). Instrument-wise, Bass improves by 5.3--6.1 points in the four non-classical subsets, suggesting that the additional acoustic coverage is particularly useful for some instrument classes.

Our offline-only model is competitive with YourMT3+ on MAPS and MulTTiPop despite replacing its specialized Perceiver TF encoder with a standard Transformer encoder, suggesting that Perceiver TF is not essential for competitive transfer. Compared with Muscriptor-Large, our model remains competitive on MAPS and RWC Prog but trails on MulTTiPop and RWC Agn. This is probably due to a substantial difference in training scale, as Muscriptor-Large is pretrained on roughly 11,000 hours of audio.

\subsection{Analysis and Limitations}
The gains from online generation become modest once diverse offline supervision is available and remain domain dependent. In addition, NSynth provides only coarse instrument labels, which limits program-level supervision, while Pianoteq is biased toward keyboard instruments and does not provide comparable coverage across the full instrument set. In addition, common post-processing such as mastering are not included. Thus, the proposed method should be viewed primarily as a way to study and extend synthetic supervision rather than as a replacement for diverse paired recordings.


\section{Conclusion}
We introduced an online sampler--renderer framework that separates symbolic structure from timbral realization in synthetic AMT data, together with a simplified multi-channel architecture that attains competitive results without a specialized encoder.
Our experiments show that score-level structure can be moderately relaxed while retaining useful supervision, whereas broader timbral coverage consistently improves transfer. Online-generated data also complement offline training, although the gains become modest and domain dependent in the strongest setting. These results suggest prioritizing coverage of note-level attributes and their acoustic realizations over reproducing realistic complete scores when designing synthetic AMT data.


\clearpage

\section{ACKNOWLEDGMENT}
AI tools (Chatgpt 5.6 sol) were used for language polishing and did not con-
tribute to the technical content of this work.
\section{COMPLIANCE WITH ETHICAL STANDARDS}
This is a computational study using existing datasets and
synthetically generated audio, for which no ethical approval
was required.

\bibliographystyle{IEEEbib}
\bibliography{references}
\end{document}